# Bending strain induced thermal conductivity suppression in freestanding $BaTiO_3$ and $SrTiO_3$ membranes

*Ziyan Qian* [a, ‡], *Guangwu Zhang* [a, ‡], *Weikun Zhou* [b], *Tsukasa Katayama* [b] *, *Qiye Zheng* [a,] *

[a] Department of Mechanical and Aerospace Engineering, The Hong Kong University of Science and Technology, Hong Kong SAR, Hong Kong

[b] Research Institute for Electronic Science, Hokkaido University, N20W10, Kita, Sapporo 001-0020, Japan

* Corresponding authors.

Email addresses: katayama@es.hokudai.ac.jp (T. Katayama), qiyezheng@ust.hk (Q. Zheng)

## Abstract

Freestanding perovskite oxide membranes provide a novel platform for "elastic strain engineering", enabling the manipulation of phonon transport free from substrate clamping. In this work, we investigate the thermal transport properties of strontium titanate ($SrTiO_3$) and barium titanate ($BaTiO_3$) membranes subjected to self-formed crease induced inhomogeneous strain. By integrating spatially resolved Frequency-Domain Thermoreflectance (FDTR) with micro-Raman spectroscopy, we observe a sharp, localized suppression of thermal conductivity ($k$) in high-curvature regions. Specifically, $k$ is reduced from 4.43 to 3.62 W/(m·K) in $SrTiO_3$ and from 2.27 to 1.81 W/(m·K) in $BaTiO_3$ at the crease centers, directly correlating with the local strain distribution. First-principles calculations reveal that, unlike uniform strain, the symmetry breaking induced by strain gradients significantly broadens phonon dispersion and enhances scattering rates. These findings not only elucidate the microscopic mechanisms governing phonon-strain coupling but also demonstrate the potential of inhomogeneous strain fields as a potent tool for designing dynamic solid-state thermal switches and active thermal management devices.

## 1 Introduction

The relentless miniaturization and three-dimensional integration of microelectronic devices have pushed power densities to unprecedented levels, making thermal management a critical bottleneck in the "post-Moore" era[1]. In modern data centers, cooling systems account for nearly 40% of total electricity consumption, highlighting an urgent need for energy-efficient thermal solutions[2]. Beyond passive heat spreading, there is a growing demand for active thermal management technologies—such as thermal switches and regulators—that can dynamically modulate heat flow in response to external stimuli[3]. These devices are essential for optimizing chip performance, protecting sensitive components from thermal shock, and enabling advanced solid-state refrigeration cycles.

To date, various mechanisms have been explored to realize solid-state thermal switching, including temperature-induced phase transitions, electrochemical intercalation, and magnetic field modulation. However, these approaches face significant practical limitations. Phase-change materials (e.g., $VO_2$, $Ge_2Sb_2Te_5$) operate only at fixed transition temperatures and often suffer from hysteresis[4,5]; electrochemical methods rely on slow ion migration and liquid electrolytes, making them incompatible with fast-response electronic integration[6–10]; and magnetic switching typically requires cryogenic temperatures or prohibitively large magnetic fields[11–14]. Consequently, manipulating phonon transport via mechanical strain in solid-state dielectrics offers a more promising pathway for integrated, high-frequency thermal management. Strain engineering effectively modulates thermal transport properties by altering lattice distortion, crystal phases, and domain structures, thereby influencing phonon scattering mechanisms. Rivadulla *et al.* utilized substrate lattice mismatch to introduce static strain in $PbTiO_3$ (PTO) thin films; they found that a 1.5% epitaxial compressive strain increased $k$ by 61% through the modulation of domain wall density and type[15]. Subsequently, the same team demonstrated that a 1.15% tensile strain in epitaxial $SrTiO_3$ (STO) thin films reduced $k$ by 60%, confirming the critical role of strain gradients in regulating phonon scattering within ferroelectric phases[16]. Significant contributions have also been made by domestic researchers. Ning *et al.* reported a 33% difference in $k$ between R-like and T-like phases of $BiFeO_3$ (BFO) films induced by substrate static stress, noting that growth conditions and domain wall density had a comparatively minor impact[17]. Moving towards dynamic control, Zeng *et al.* achieved a dynamic thermal switch ratio of 1.42 in polycrystalline graphite films via 1% uniaxial tension, revealing the mechanism of mechanical strain on phonon-grain boundary scattering[18]. Nie *et al.* observed a tenfold irreversible increase in interfacial thermal resistance between the Al transducer layer and a 5 nm freestanding BFO film under 3.5% uniaxial tensile strain, attributing this to strain-induced polarization rotation[19]. More recently, Yang *et al.*, using thermal bridge measurements and electron energy loss spectroscopy, revealed the regulation of phonon band structures by inhomogeneous stress in Si nanowires; notably, a strain gradient of 0.112%/nm reduced $k$ by 34%, presenting a distinct contrast to results under uniform strain[20]. However, traditional static strain engineering remains limited by substrate clamping, making it difficult to achieve dynamic, reversible control at the device level.

To overcome the limitations of substrate clamping, freestanding single-crystalline perovskite oxide membranes have recently emerged as a novel platform. Liberated from the substrate, these membranes exhibit super-elasticity and can sustain large, reversible mechanical deformations (up to ~8%) without fracture[21–23]. This capability opens new avenues for exploring "elastic strain engineering" where thermal transport can be dynamically tuned via bending or stretching. Based on this, researchers have begun to rationally engineer advanced functionalities in ferroelectrics, ranging from enhanced polarization and flexible memory to high-performance sensors[24–26]. Despite extensive investigations into strain-engineered electronic and ferroelectric properties, the intricate phonon transport mechanism subjected to large, non-uniform and reversible strain fields remains largely unexplored[27–29]. This is especially frustrating given that efficient heat dissipation and active thermal management—such as solid-state thermal switching—stand as critical bottlenecks to the performance and lifespan of next-generation microelectronics. Perovskite ferroelectric oxides ($ABO_3$), such as $SrTiO_3$ and $BaTiO_3$, are particularly attractive candidates for this purpose due to the strong coupling between their lattice dynamics, polarization, and external fields[30,31]. Previous studies have demonstrated that strain can significantly alter the $k$ of these materials by modifying phonon dispersion relations and scattering rates[32,33]. However, although modulations in $k$ have been observed, the effects of strain within these films are experimentally difficult to decouple from the substrate clamping effects and interfacial phonon scattering. The major challenges in experimentally quantifying the effects of inhomogeneous strain on thermal transport include applying stress exclusively without the interference of a substrate and combining thermal measurements with spatially resolved characterization of the crystal lattice and domain structure. Similarly, while ferroelastic domain walls can scatter phonons, it remains a formidable challenge to isolate their effects from the substrate-induced stress fields, which could also impede thermal transport via an increase in vibrational anharmonicity or symmetrical breaking. Consequently, questions regarding the dynamic tunability of heat flow in these functional oxides have lingered unanswered.

Here, we induce inhomogeneous strain through self-formed crease in freestanding $SrTiO_3$ and $BaTiO_3$ membranes on a PET substrate and measure its effect on thermal transport while characterizing the local lattice and domain structure using spatially resolved FDTR. Our results show that the spontaneous wrinkle-induced inhomogeneous strain gradient leads to a significant reduction in $k$, distinct from the behavior observed in bulk or clamped films. Taking advantage of the high spatial resolution of our optical thermal metrology, we directly correlate the local thermal transport properties with the strain distribution and determined by Raman spectroscopy. Coupled with first-principles Boltzmann Transport Equation (BTE) modelling, we demonstrate that the lattice distortion and symmetry breaking induced by the strain field enhance phonon scattering rates, ultimately contributing to a suppressed *k*.

## 2 Results and Discussion

Heterostructures comprising amorphous $Al_2O_3$/$ABO_3$/$Sr_3Al_2O_6$ were fabricated on $SrTiO_3$ (00*l*) single-crystal substrates ($STO_{sub}$) via pulsed laser deposition (PLD). In this architecture, the perovskite $ABO_3$ layer refers specifically to either $SrTiO_3$ (STO) or $BaTiO_3$ (BTO), where A (Ba and Sr in this work) is an alkaline earth or rare earth metal and B (Ti in this work) is a transition metal. The water-soluble $Sr_3Al_2O_6$ (SAO) film functioned as the sacrificial layer, while the $Al_2O_3$ overlayer served as a capping layer to provide mechanical support and prevent rupture in the perovskite films. Fig.1 (a-c) illustrates the fabrication and transfer process of the freestanding $ABO_3$/$Al_2O_3$ films. First, the water-soluble SAO sacrificial layer, the target $ABO_3$ film, and the $Al_2O_3$ capping layer were epitaxially grown in sequence on an $STO_{sub}$ (Fig. 1a). Subsequently, the heterostructure was immersed in deionized water. The selective dissolution of the SAO sacrificial layer caused the upper $Al_2O_3$/$ABO_3$ bilayer to detach from the substrate and float on the water surface (Fig. 1b). Crucially, the floating membrane was manually inverted using tweezer in the water to reverse the stacking order. Finally, the membrane was scooped up using a PET substrate, resulting in the final $ABO_3$/$Al_2O_3$/PET assembly (Fig. 1c). Following this transfer, the assembly was air-dried at room temperature for 5 h to remove residual moisture.

Fig. 1(d) and (e) present cross-sectional scanning electron microscopy (SEM) images of the as-grown $Al_2O_3$/$ABO_3$/SAO/$STO_{sub}$ heterostructures, respectively. These images reveal distinct interfaces and uniform layer thicknesses, measuring approximately 220 nm for STO, 340 nm for BTO, and 700 nm for $Al_2O_3$. Notably, the morphological differences between the STO and BTO films are directly correlated with their distinct strain states. As shown in Fig. 1(d), the STO film exhibits a remarkably smooth interface, whereas the BTO film in Fig. 1(e) displays significant roughness. This contrast is further elucidated by the X-ray diffraction (XRD) patterns shown in Fig. 1(f), which compares the as-grown and transferred films. Both samples exhibit high crystallinity with a strong *c*-axis orientation, evidenced by sharp (00*l*) diffraction peaks. For the STO film, the smooth morphology suggests a coherent growth mode facilitated by the close lattice matching with the SAO buffer. This minimizes strain accumulation, a conclusion corroborated by the negligible shift in the STO diffraction peak before and after transfer, indicating a nearly stress-free state[34]. Conversely, the roughness observed in the BTO film is attributed to the accumulation of strain energy arising from the substantial lattice mismatch with the substrate. This morphological instability is indicative of the high elastic energy stored during heteroepitaxial growth, which is subsequently released upon transfer[35]. As highlighted by the shaded region in Fig. 1(f), the removal of the SAO sacrificial layer releases the in-plane compressive strain imposed on the BTO film. This relaxation leads to an expansion of the in-plane lattice constants ($a$ and $b$) and a simultaneous contraction of the out-of-plane lattice constant ($c$) via the Poisson effect[36,37]. Consequently, the (00$l$) peaks of the transferred BTO/$Al_2O_3$/PET structure shift towards a higher $2\theta$ angle.

Following the transfer process, the SEM images of the transferred STO (Fig. 1g) and BTO (Fig. 1h) films reveal the spontaneous formation of creases and cracks along the surface. These macroscopic defects suggest that the compressive strain accumulated in the oxide thin films was effectively released upon transfer to the compliant PET substrate. The formation of such buckling patterns or fractures is a typical strain relaxation mechanism for freestanding oxide membranes released from their growth substrates.

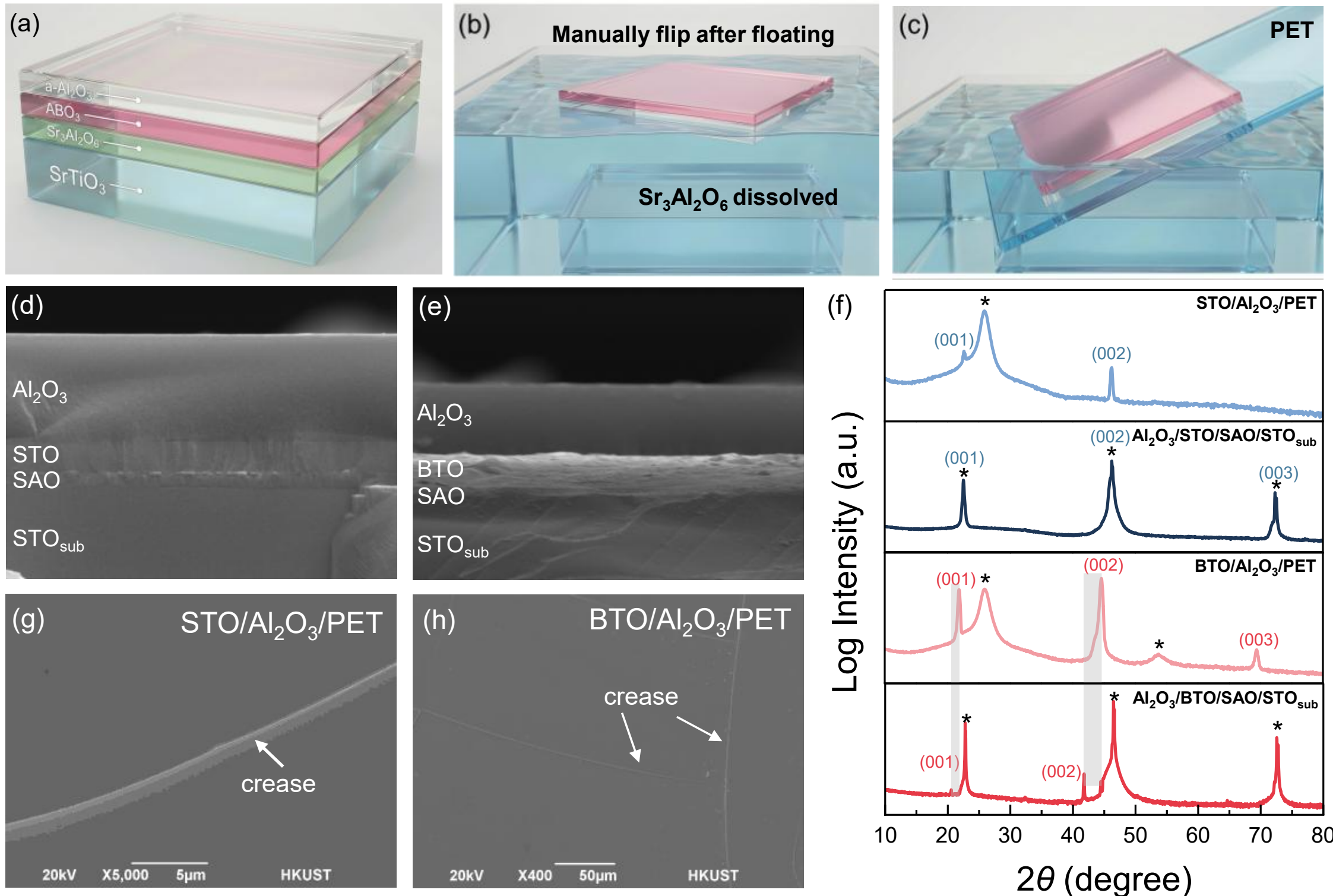

**Figure 1. Fabrication of freestanding PLD membrane (BTO & STO) by chemical lift-off.** (a) The multilayer structure grown on an STO substrate, consisting of an SAO sacrificial layer, the $ABO_3$ film, and an amorphous $Al_2O_3$ capping layer. (b) Release of the $ABO_3$/$Al_2O_3$ film floating on water after dissolving the SAO layer. (c) Transfer of the released film onto a PET substrate. Cross-section SEM image of (d)

$Al_2O_3$/STO/SAO/$STO_{sub}$ and (e) $Al_2O_3$/BTO/SAO/$STO_{sub}$. (f) XRD curves of as grown and free-standing membranes. Substrate peaks, arising from either PET or the $STO_{sub}$, are marked with asterisks (*). Surface morphology of (g) STO/$Al_2O_3$/PET and (h) BTO/$Al_2O_3$/PET. Top-view SEM images of the transferred (g) STO and (h) BTO films, respectively. The images reveal the spontaneous formation of creases and cracks along the film surfaces, indicating strain relaxation during the transfer process.

To further investigate the topological nature of the surface creases observed in the SEM images and elucidate the strain distribution induced by the morphological instabilities during the transfer process, we performed surface profilometry and micro-Raman spectroscopy. Fig. 2(a) and (b) present the cross-sectional height profiles of the creases in STO and BTO films, respectively. Interestingly, the scans reveal distinct geometric characteristics for the two materials. The crease on the STO film (Fig. 2(a)) manifests as a significant upward protrusion (or ridge) with a peak height of approximately 3 μm. In sharp contrast, the feature on the BTO film (Fig. 2(b)) appears as a downward depression (or trench) extending about 1.5 μm below the baseline. This morphological dichotomy highlights the complex interplay between residual film stress and the flexible PET substrate during the transfer process. In addition, the two distinct deformation modes result in non-uniform strain distributions across the film thickness, as illustrated in the insets. For the STO film, which forms a ridge-like convex structure, the deformation leads to a state where the top surface is under compression (blue arrows) while the bottom surface undergoes tension (red arrows). In contrast, the BTO film develops a valley-like concave morphology, resulting in an inverted strain gradient with tensile strain at the top surface and compressive strain at the bottom interface.

To experimentally validate this strain analysis, spatially resolved Raman spectroscopy was conducted on the BTO sample, focusing on the $B_1$, $E$(LO+TO) phonon modes which are sensitive to lattice distortions. As shown in Fig. 2(c), spectra were collected from the center of the crease (P1) extending to the relaxed flat region (P5). A distinct frequency upshift (blue shift) of the peak position is observed, moving from 318 $cm^{-1}$ at the stress-free region (P5) to 315 $cm^{-1}$ at the crease center (P1). This hardening of the phonon mode is a hallmark of tensile stress, providing strong spectroscopic evidence that aligns perfectly with the concave geometry observed in the profilometer data. These localized, non-uniform strain fields are expected to play a pivotal role in modifying the phonon mean free path, thereby offering a mechanism to tune the $k$ of the perovskite films.

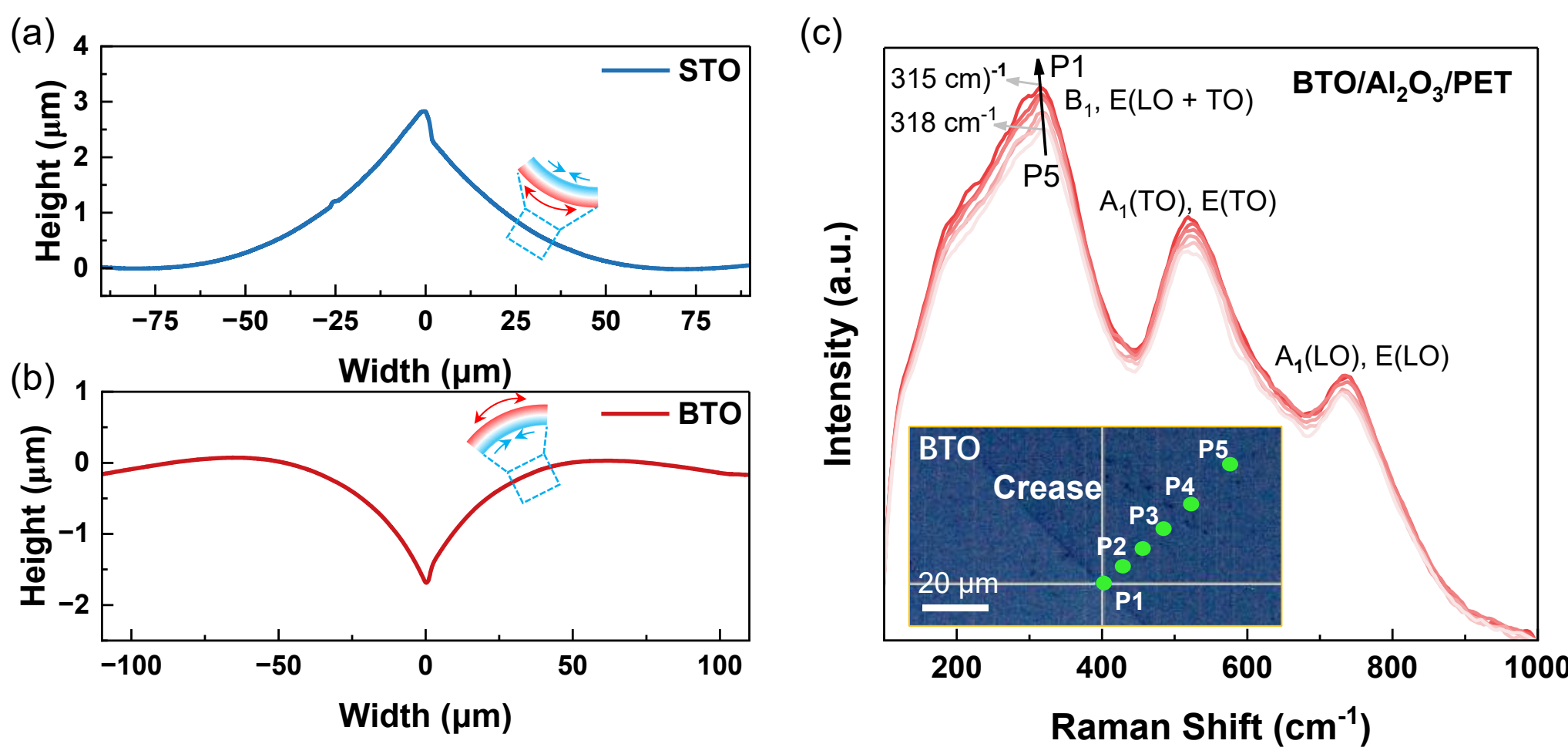

**Figure 2. Morphological characterization and local strain analysis of the transferred perovskite films.** (a, b) Surface height profiles of the STO and BTO films measured across the creases, respectively. The insets schematically illustrate the crease-induced strain states, where red and blue arrows denote tensile and compressive strains, respectively. Note that the STO film exhibits a ridge-like (convex) morphology, whereas the BTO film displays a valley-like (concave) deformation, leading to distinct local strain environments at the film surface. (c) Spatially resolved Raman spectra of the BTO/$Al_2O_3$/PET film collected at various positions near a crease. The inset shows the corresponding optical micrograph, with points P1 to P5 indicating the laser spot locations from the crease center to the flat region. A noticeable blue shift of the $B_1$, $E$(LO+TO) mode from 315 $cm^{-1}$ (flat region, P5) to 318 $cm^{-1}$ (crease, P1) confirms the presence of localized compressive strain, consistent with the concave morphology observed in (b).

We employed high-resolution FDTR to investigate the impact of microscopic morphology on thermal transport properties within the crease regions. Before FDTR measurements, a 100 nm Au layer was sputtered onto the sample as a transducer, ensuring efficient photothermal conversion and temperature sensing. As illustrated in Fig. 3(a), the sample consists of an Au/$ABO_3$/$Al_2O_3$PET stack. FDTR operates by using two continuous-wave (CW) laser beams: a modulated pump beam and a probe beam. The pump beam heats the sample at varying modulation frequencies, creating a periodic temperature oscillation. The probe beam, reflected from the sample, detects this temperature response via the thermoreflectance effect, where the surface reflectivity changes linearly with small temperature variations[38]. By measuring the phase lag between the pump and probe signals across a range of frequencies and fitting the data to a 3D heat diffusion model, we can extract the sample's thermal properties[39].

Prior to measurement, we performed a rigorous sensitivity analysis to quantify the impact of individual thermal parameters within the $Au/BTO/Al_2O_3/PET$ multilayer stack on the FDTR measurement to optimize the measurement conditions and increase accuracy. The measurement sensitivity $S_\gamma$ to a specific parameter $\gamma$ is defined as $S_\gamma = \partial\phi/\partial \ln\gamma$, with $\phi$ denoting the FDTR phase lag[39]. A higher $S_\gamma$ value indicates greater accuracy in determining $\gamma$.

Fig. 3(b) illustrates the phase sensitivity to various thermal parameters from 6 to 50 MHz. The analysis reveals that the measurement is largely insensitive to the supporting layers and bottom interfaces ($G_{Al2O3/PET}$, $G_{BTO/Al2O3}$, and $k_{Al2O3}$), thereby simplifying the thermal model. The dominant contribution comes from the BTO layer thermal conductivity ($k_{BTO}$), which maintains high sensitivity across the full spectrum. To account for the minor yet non-negligible influence of the top interface, we performed a multiparameter fit extracting both $k_{BTO}$ and Au/BTO interface conductance $G_{Au/BTO}$. This approach effectively decouples the interface resistance from the intrinsic film properties, ensuring a precise determination of $k_{BTO}$.

To elucidate the microscopic impact of local deformation on heat transport, we performed spatially resolved thermal conductivity measurements across typical creases in freestanding STO and BTO films. Fig. 2(c) and (d) display representative frequency-dependent phase signals collected at various locations relative to the crease (as shown in the insets). The experimental data (open circles) exhibit excellent agreement with the theoretical predictions (solid lines) derived from the multilayer heat diffusion model over the entire frequency range (6–50 MHz). To further validate the measurement precision, we plotted sensitivity bounds (dashed lines) corresponding to a ±15% variation in the extracted cross-plane thermal conductivity. The fact that the experimental data are tightly confined to the best-fit curves—and clearly distinguishable from the ±15% deviation lines—demonstrates the high sensitivity of our FDTR setup and the reliability of the extracted $k$ values, even in the vicinity of the structural crease.

Fig. 3(e) and (f) display the local thermal conductivity ($k$) profiles superimposed onto the height topography of the deformed films. To establish a precise correspondence between geometric deformation and thermal transport, specific measurement spots are annotated directly on the surface profile curves (P1–P8 for STO; P1–P6 for BTO). In the convexly bent STO film (Fig. 3(e)), the thermal conductivity reveals a pronounced spatial dependence governed by local curvature. While $k$ remains high at approximately 4.43 W/(m·K) in the quasi-flat peripheral regions (e.g., P1, P8), it exhibits a sharp reduction to 3.62 W/(m·K) near the apex of the crease (P4, P5), where the tensile strain is maximized. Intriguingly, the concave, valley-like crease in the BTO film (Fig. 3(f)) exhibits a strictly analogous behavior despite the opposing curvature. Here, $k$ reaches a minimum of 1.81 W/(m·K) precisely at the valley floor (P3)—significantly lower than the 2.27 W/(m·K) recorded at the flatter edges.

This spatial mapping confirms that thermal suppression is strictly confined to regions of high local curvature, regardless of whether the macroscopic deformation is convex or concave. We attribute this localized reduction to the complex interplay between strain fields and phonon dynamics. As detailed in our first-principles calculations (Fig. 4), the two materials exhibit distinct intrinsic responses: BTO displays a monotonic variation in $k$ across compression and tension, whereas STO exhibits a symmetric reduction in $k$ under both strain states. Consequently, for STO, the reduction is intrinsic to the lattice distortion regardless of the strain sign. However, the universal suppression observed in both films—even where bond stiffening might be expected—suggests that extrinsic factors are dominant. Specifically, the steep strain gradients within the crease likely induce anharmonicity and potential microstructural defects, acting as potent scattering centers for phonons and thereby impeding thermal transport across the deformed region.

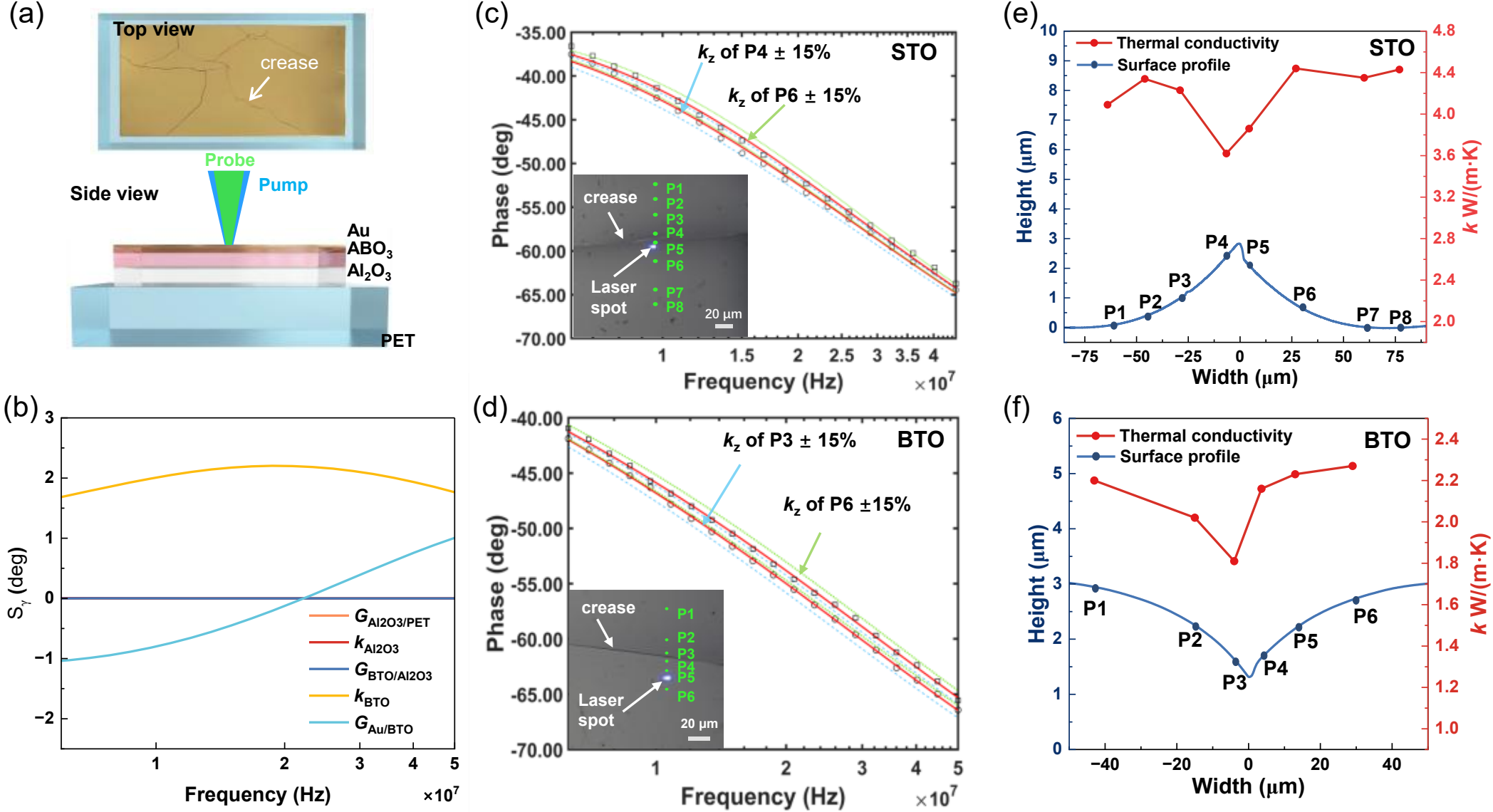

**Figure 3. FDTR measurement setup and thermal conductivity characterization across film creases.** (a) Schematic illustration of the FDTR experimental setup. The top view shows crease on the transferred film. The side view depicts the sample cross-section, where a 100 nm sputtered Au layer serves as the transducer on top of the $ABO_3/Al_2O_3$/PET stack. (b) Phase sensitivity ($S_\gamma$) analysis of the Au/BTO/$Al_2O_3$/PET structure as a function of modulation frequency, identifying the sensitivity of the signal to various thermal parameters (thermal conductivity ($k$) and interface thermal conductance ($G$)). (c, d) Representative FDTR phase signals versus modulation frequency for (c) STO and (d) BTO free-standing films. The open circles denote

experimental data, while the solid lines represent the best theoretical fits used to extract $k$. The insets display optical micrographs of the measurement areas, mapping the spatial distribution of the probe spots (P1–P8 for STO; P1–P6 for BTO) across the bending crease. To demonstrate measurement sensitivity, the dashed lines indicate the calculated phase deviations corresponding to a ± 15% variation in $k$ for selected points near (e.g., P4 in STO, P3 in BTO) and away from (P6) the crease center. (e, f) Spatially resolved $k$ (red spheres, right axis) and corresponding surface height profiles (blue curves, left axis) measured across a convex crease in the STO film (e) and a concave crease in the BTO film (f). The labeled points (P1–P8 in (e); P1–P6 in (f)) indicate the specific positions of the FDTR measurement spots relative to the crease morphology, revealing a reduction in thermal conductivity at the highly deformed regions.

To gain microscopic insight into how strain fields regulate phonon transport in our freestanding perovskite membranes, we carried out first-principles BTE calculations for STO and BTO under both uniform and inhomogeneous strains, as summarized in Fig. 4 and 5. Fig. 4(a) shows that for STO, the thermal conductivities along the three principal axes all decrease monotonically with the increasing tensile strain $\varepsilon_{xx}$, Under compressive strain, $k_a$ and $k_c$ first increase and then decrease, while $k_b$ consistently decreases. This behavior indicates that in STO the dominant role of strain is to enhance anharmonic phonon-phonon scattering and shorten phonon mean free paths. Quantitatively, our calculations show that applying ±2% uniaxial strain along the $x$ direction reduces the total thermal conductivity of STO by ~16-22%, corresponding to a modest thermal switching ratio of ~1.2-1.3. In contrast, BTO exhibits a much richer and more anisotropic response due to its ferroelectric nature. As shown in Fig. 4(c), under $\varepsilon_{xx}$, the thermal conductivity of BTO increases with increasing compressive strain, whereas under tension it generally decreases as $\varepsilon_{xx}$ becomes more positive. Strikingly, within the tensile range between approximately 1% and 2%, the thermal conductivities along the $b$ and $c$ axes show an upturn: $k_b$ and $k_c$ increase in this interval even though the overall trend under tension is a reduction of $k$. This anomalous enhancement along the $b$ and $c$ directions is closely linked to a strain-induced reorientation of ferroelectric polarization. Previous first-principles studies have shown that in perovskite ferroelectrics, strain can drive transitions between different ferroelectric phases and rotate the polarization direction, thereby modifying the symmetry, soft-mode character, and domain structure[40–42]. When $\varepsilon_{xx}$ is in the range of 1-2% tension, the polarization in BTO tends to reorient away from its original direction toward in-plane or tilted configurations, which in turn alters the phonon dispersion and scattering landscape in an anisotropic manner. Notably, under the same ±2% uniaxial strain along $x$ direction, BTO exhibits a much stronger suppression of thermal conductivity, reaching ~63–68% (a thermal switching ratio up to ~3). This pronounced reduction in BTO arises from a strain-driven structural distortion from the tetragonal phase toward an orthorhombic-like phase, where the associated symmetry breaking markedly enhances phonon scattering rates. The distinct strain dependences of STO and BTO thereby highlight the strong coupling between lattice dynamics, ferroelectric polarization, and thermal transport in ferroelectric perovskites.

These trends can be directly related to the strain-dependent phonon dispersions shown in Fig. 4 (b) and (d). For both STO and BTO, compressive strain leads to an overall hardening of the phonon branches: the dispersions shift to higher frequencies and certain acoustic and optical branches become steeper, indicating increased phonon group velocities in parts of the Brillouin zone. In contrast, tensile strain causes an overall softening, with several branches—especially low-frequency modes—shifting to lower frequencies and becoming more densely packed. These hardening and softening are accompanied by a pronounced broadening of the dispersion curves, in the sense that more phonon branches lie in close proximity in energy and momentum space. As discussed by Yang et al's work[20], such broadening of the dispersion under strain implies that a larger number of phonon modes can simultaneously satisfy the energy and momentum conservation conditions required for three-phonon and four-phonon scattering (including Umklapp processes). Consequently, the available scattering phase space is significantly expanded, which in turn increases phonon scattering rates and further suppresses the thermal conductivity.

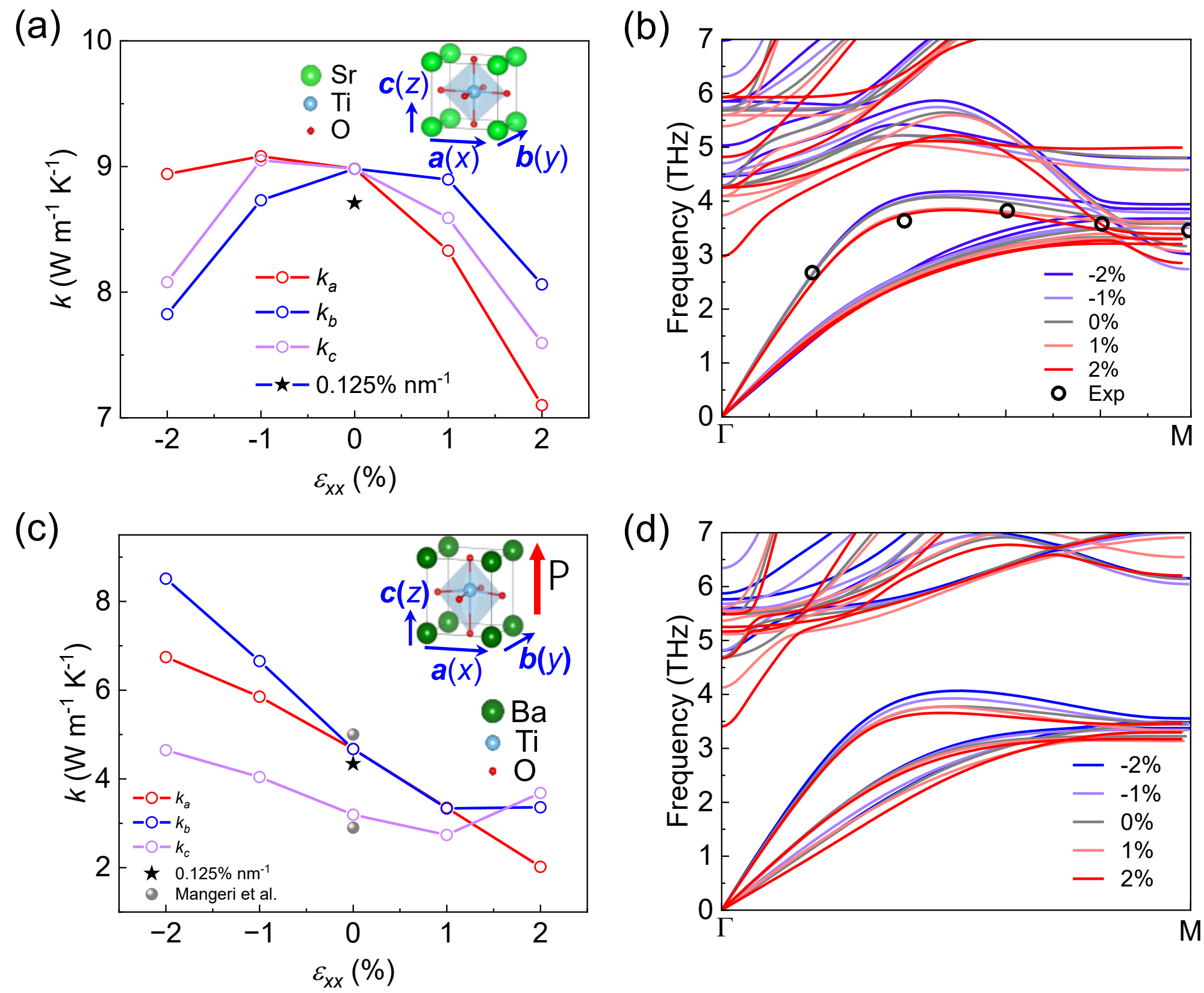

**Figure 4.** First principles BTE calculated thermal conductivities (left) for bulk STO (a,b) and BTO (c,d) and dispersion (right) curves under varied uniform strain along the *x*-axis. The insets in (a) and (c) show the corresponding crystal structure and polarizations.

This calculated mode-resolved scattering rates under uniform and inhomogeneous strains are shown in Fig. 5. The thermal conductivity of STO under free strain is 8.98 W/(m·K). The out-of-plane thermal conductivity of BTO is 3.1 W/(m·K), while the in-plane thermal conductivity is 4.67 W/(m·K) (close to experimental values of 5 W/(m·K) and 2.9 W/(m·K)[43]). Under purely uniform strain (−2% to +2%), STO exhibits only modest changes in the overall phonon scattering rates with increasing strain magnitude, indicating that the variation of thermal conductivity in STO cannot be solely attributed to enhanced anharmonic scattering, but is likely strongly influenced by strain-induced modifications of the phonon dispersions, such as band mixing, branch crossings, and subtle changes in group velocities. In contrast, BTO shows a pronounced increase in phonon scattering rates under both compressive and tensile strains, implying that in BTO the evolution of thermal conductivity, especially its reduction under tension, is governed by the combined effects of dispersion softening/broadening and substantially enhanced phonon-phonon scattering. When a finite strain gradient is introduced— modeled here as inhomogeneous strain fields with gradients of 0.05% $nm^{-1}$ and 0.125% $nm^{-1}$—the scattering rates across almost the entire frequency spectrum are further elevated relative to the uniformly strained case at the same average strain. The enhancement is particularly pronounced for low-frequency acoustic phonons, which are the main heat carriers. Physically, an inhomogeneous strain field acts as a slowly varying elastic disorder potential that breaks translational invariance and disrupts phonon coherence; it can be viewed as a continuous distribution of "virtual interfaces" or domain-wall-like regions that provide additional scattering channels beyond intrinsic anharmonic processes. In our BTE framework, this effect is captured by explicitly introducing a strain-gradient-induced scattering contribution, $\tau_{sg}^{-1}$ (see details in method). Our preliminary work shows that an out-of-plane strain gradient of 0.125% $nm^{-1}$ can produce an additional ~7% reduction in the in-plane thermal conductivity in both STO and BTO (Fig.4). Following the interpretation of Yang et al's work[20], the dispersion broadening and mode mixing induced by such non-uniform strain dramatically increase the number of phonon pairs and triplets that fulfill the scattering selection rules, thereby amplifying the total scattering rate and driving the thermal conductivity to even lower values than in the uniformly strained case. Taken together, the BTE calculations reveal a consistent microscopic picture: in STO, both compressive and tensile uniform strains reduce $k$ along all three axes primarily through complex strain-induced changes in the phonon dispersion, with only relatively minor contributions from changes in scattering rates; in BTO, uniform compression hardens the spectrum and, despite increased scattering rates, can maintain or even enhance $k$ along some directions, whereas tensile strain softens and broadens the phonon dispersions and, together with the strongly increased scattering rates, generally lowers $k$, with a notable 1-2% tensile window where polarization reorientation leads to a relative enhancement of $k$ along the $b$ and $c$ axes. In both materials, the introduction of inhomogeneous strain gradients significantly increases phonon scattering rates beyond the uniform-strain limit, providing a natural explanation for the experimentally observed strong suppression of thermal conductivity localized at the highly curved, strongly bent regions of the freestanding membranes.

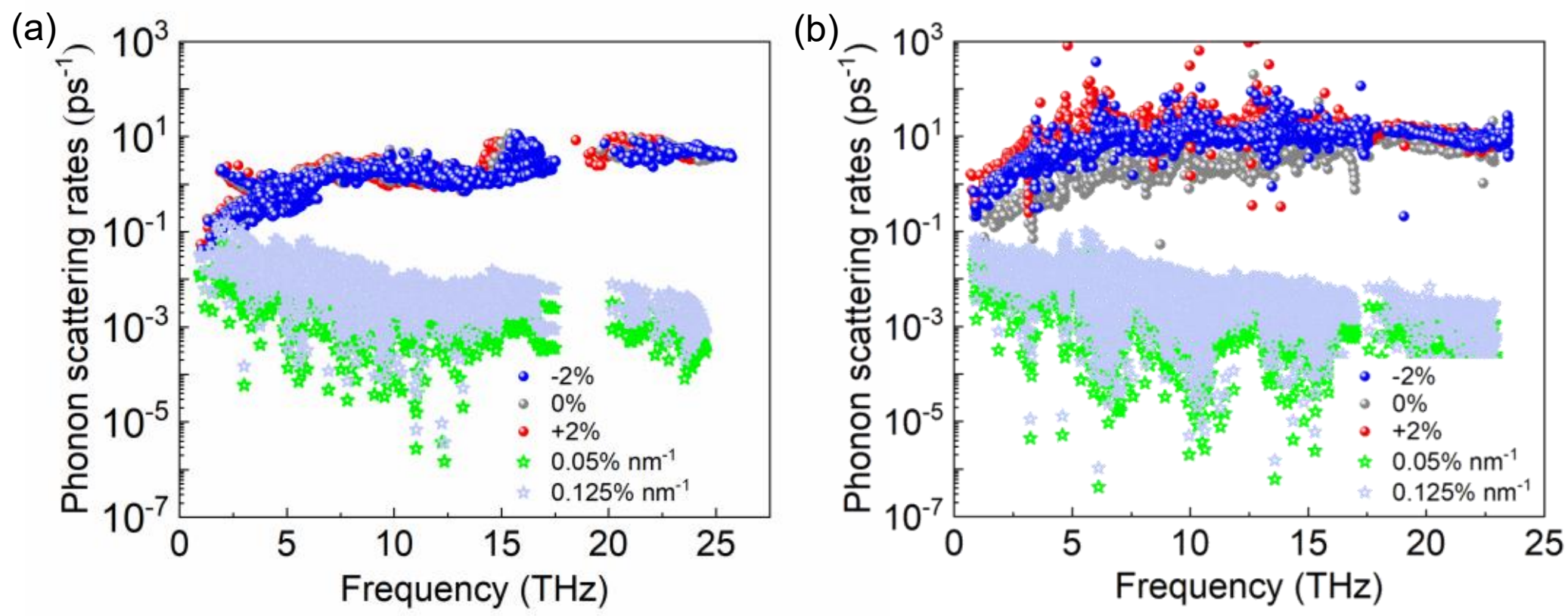

**Figure 5. Scattering rates under uniform and inhomogeneous strains**. (a) Scattering rates of STO under uniform (ranging from -2% to 2%) and inhomogeneous strains (0.05% nm$^{-1}$, 0.125% nm$^{-1}$). (b) Scattering rates of BTO under uniform (ranging from -2% to 2%) and inhomogeneous strains (0.05% nm$^{-1}$, 0.125% nm$^{-1}$).

## 3. Conclusion

This study establishes a direct link between macroscopic mechanical deformation and microscopic phonon transport in freestanding perovskite ferroelectrics, overcoming the longstanding limitations of substrate clamping. By fabricating high-quality $SrTiO_3$ and $BaTiO_3$ membranes, we successfully induced large, reversible strain gradients via self-formed crease and mapped their impact on heat flow with high spatial resolution. Our results demonstrate that inhomogeneous strain fields are far more effective at suppressing thermal conductivity than uniform strain. The experimental observation of reduced thermal transport in highly curved regions is quantitatively supported by first-principles modeling, which identifies strain gradient-induced symmetry breaking as a critical factor in enhancing phonon scattering rates. Furthermore, while $SrTiO_3$ behavior is dominated by dispersion modification, $BaTiO_3$ exhibits a complex response driven by the interplay of ferroelectric polarization and soft-mode hardening. Ultimately, this work validates "elastic strain engineering" as a viable strategy for active thermal regulation. By decoupling intrinsic material properties from substrate effects, we pave the way for the development of mechanically tunable thermal logic devices and efficient cooling solutions essential for the post-Moore era of microelectronics.

## Acknowledgements

We would like to acknowledge the financial support from the Nano & Material Technology Development Program through the National Research Foundation of Korea (NRF) funded by the Ministry of Science and ICT (RS-2024-00444574), the National

Key R&D Program of China (2024YFB4405700), the HKUST-HKUST(GZ) Cross-campus Collaborative Research Scheme under the “1+1+1” Joint Funding Program (No. 2025A0505000029), the Innovation and Technology Fund (ITF) from the Hong Kong Innovation and Technology Commission under (ITS/135/23), and the Guangdong Basic and Applied Basic Research Foundation (2025A1515012898). We also thank the HKUST Fok Ying Tung Research Institute and the National Supercomputing Center in Guangzhou Nansha Sub-center for providing high-performance computational resources, and acknowledge funding from the Frontier Technology Research for Joint Institutes with Industry Scheme, sponsored by the Center on Smart Sensors and Environmental Technologies at HKUST.

## 4 Methods

### 4.1 Experiments

Heterostructures comprising $Al_2O_3$/STO (or BTO)/SAO were fabricated on $SrTiO_3$ (001) single-crystal substrates via PLD. The deposition conditions were optimized for each layer: the STO and BTO layers were grown at a substrate temperature of 790 °C under an oxygen partial pressure $6 \times 10^{-3}$ Torr; the SAO layer was deposited at 850 °C and $7.5 \times 10^{-6}$ Torr; and the $Al_2O_3$ capping layer was deposited at 30 °C and $7.5 \times 10^{-6}$ Torr. Layer thicknesses were characterized using scanning electron microscopy (SEM; JSM-7100F, JEOL Ltd., Tokyo, Japan). The thicknesses of STO, BTO, and $Al_2O_3$ were measured to be 220, 340, and 710 nm, respectively. To detach the functional layers, the as-grown samples were immersed in deionized water for 24 h to ensure the complete dissolution of the SAO sacrificial layer. Subsequently, the $Al_2O_3$/STO (or BTO) layer was flipped by using a tweezer, and a 0.3 mm-thick PET support was employed to lift off the floating $Al_2O_3$/STO (or BTO) membranes. Finally, the transferred films were dried at room temperature to eliminate residual moisture and enhance adhesion between the film and the PET substrate.

Scanning electron microscopy (SEM; JSM-7100F (JEOL Ltd., Tokyo, Japan)) was used to characterize the thickness and the morphology of the films. The crystal structures of the films were characterized by XRD with using a Panalytical X'Pert Pro diffractometer in a $\theta$–$2\theta$ and configuration with Cu K$\alpha$ radiation ($\lambda$ = 1.5405 Å). The height profiles of the creases in STO and BTO freestading films was measured by an Alpha-Step D300 profiler (KLA-Tencor, California, USA) at a rate of 0.05 mm/s.

Raman spectroscopy was used to characterize the strain ratio. All measurements were performed at room temperature using a Renishaw RM 3000 Micro-Raman system (InVia, Renishaw Co., UK) employing 325 nm laser that was focused on a diffraction limited spot by a 50× laser. The Raman scattered light was dispersed using a grating with 2400 g/mm, and the laser power was set to 2 mW to limit the steady state heating of the films below 5 K.

The FDTR system employs a 445 nm pump laser modulated in 5 to 60 MHz, and a CW 532 nm probe laser which are focused by a M Plan Apo 20× objective lens (Mitutoyo Ltd, Tokyo, Japan) to a $1/e^2$ laser spot ≈ 1.9 μm. In our measurements, the pump laser power is set to 3 mW, while the probe laser power is set to 1 mW, the steady-state temperature rise is calculated to be less than 10 K. The frequency-dependent phase lag between the probe and pump lasers was captured using a lock-in amplifier (HF2LI, Zurich). The sample films were coated by a 100 ± 2 nm Au transducer layer by magnetron sputtering which absorbs the pump for periodic heating and produce thermoreflectance signal to monitor the surface temperature evolution. The thickness of the Au layer was measured by an Alpha-Step D300 profiler (KLA-Tencor, California, USA) using an Au/Si coated together with the sample. The thermal conductivity of the Au layer ($k_{Au}$) was determined by fitting the FDTR measurement data of an Au/$SiO_2$ (2 μm)/Si control sample fabricated in the same batch. This method was adopted due to the high sensitivity of the FDTR signal to $k_{Au}$ in this structure, as well as the well-established thermal properties of the $SiO_2$ and Si layers. The volumetric heat capacities of Au[44] was sourced from the literature. The fixed parameters for FDTR fitting are summarized in Table S1. The FDTR measurement error in $k$ is estimated by the error propagation method, assuming a 3% uncertainty associated with known parameters. The FDTR metrology is detailed in prior studies[38,39,45].

## 3.2 Calculations

### 3.2.1 Neuroevolution machine-learning potential construction

The geometry optimizations of STO and BTO were performed using first-principles calculations within the framework of density functional theory (DFT) as implemented in the Vienna *ab initio* simulation package (VASP)[46,47]. The exchange-correlation functional was described by Perdew-Burke-Ernzerhof revised for solids (PBEsol) form[48]. The plane-wave kinetic energy cutoff was set at 550 eV, with a total-energy convergence criterion of $1\times 10^{-7}$ eV. *ab initio* molecular dynamics (AIMD) simulations were performed in the canonical (NVT) ensemble using the Nosé thermostat to extract configurations from primitive supercells over a temperature range of 50 K to 750 K. Additionally, the initial configurations were enriched by applying uniform strain (ranging from -3% to +3%) as well as introducing random atomic displacements and lattice perturbations. The energies and interatomic forces were then evaluated using DFT self-consistent calculations. Γ-centered k-point meshes of 1 × 1 × 1 and 4 × 4 × 4 were used. The final STO and BTO dataset contained 1600 configurations. The hyperparameters used to train the final NEP model are summarized in Table 1. The training energy, force, and virial root mean square errors (RMSE) are 0.4 meV/atom, 31.2 meV/Å, and 2.8 meV/atom, respectively as shown in Fig. 6.

**Table 1. Main hyperparameters used in the training of NEP model.** $r_c^R(r_c^A)$ defines the radial (angular) cutoffs (Å), $n_{max}^R(n_{max}^A)$ denotes the Chebyshev polynomial expansion order for the radial (angular) descriptor components, $N_{neu}$ is the number of neurons in the hidden layer, $\lambda_1(\lambda_2)$ indicates the weight of the regularization term in the loss function, $N_{gen}$ specifies the maximum steps of the generation. More details

of NEP can be found in Xu et al.'s work[49].

| Parameter | Value | Parameter | Value |
|---|---|---|---|
| $r_c^R$ | 8 | $N_{neu}$ | 50 |
| $r_c^A$ | 5 | $\lambda_1$ | 0.05 |
| $n_{max}^R$ | 12 | $\lambda_2$ | 0.05 |
| $n_{max}^A$ | 8 | $N_{gen}$ | $2 \times 10^5$ |

**Figure 6. Training results of the NEP model**. (a) Evolution of the energy, force, and virial loss functions on the training and test sets during NEP training convergence. Comparisons of the NEP-predicted and DFT-calculated (b) energy, (c) force, and (d) virial in the training dataset.

### 3.2.2 Lattice dynamic calculations

Because the finite-displacement method yields harmonic force constants referenced to the 0 K equilibrium structure, it predicts imaginary phonon modes for STO and BTO (Fig. S2), indicating a dynamic instability within the 0 K harmonic approximation, which has been proved in previous works[50,51]. Therefore, we employed the temperature-dependent effective potential approach, as implemented in the hiPhive package[52], to extract finite-temperature effective harmonic and anharmonic force constants. These force constants were then used to compute the harmonic phonon properties, including the volumetric heat capacity, phonon dispersions, phonon density of states, group velocities. We performed NVT molecular dynamics simulations using NEP on the 4 × 4 × 4 supercells at different temperatures, with a time step of 1 fs and a total simulation time of 50 ps. During the simulations, 100 configurations were extracted at 500 fs intervals. Finally, harmonic force constants were extracted from these configuration data. The cutoff distances for training force constant potentials are chosen to be 6.5 Å, 4.5 Å, and 3.5 Å. Thereafter, we investigate phonon-mediated thermal transport using the unified theory that incorporates both the particle-like and wave-like (coherence) transport regimes[53,54] in the modified FourPhonon software[55,56]. The interaction cutoffs for third-order and fourth-order IFCs are set to seventh and third nearest neighbors, respectively, which are enough to ensure the convergence. And a dense $q$ mesh of 11 × 11 × 11 is used for the BTE calculations. The Born effective charge and dielectric tensor are considered to account for long-range electrostatic interactions.

### 3.2.3 Strain gradient model

Strain gradient $\eta$ induces lattice asymmetry, making strain gradient-induced scattering an intrinsic scattering mechanism. According to Matthiessen's rule, the total phonon scattering rate ($\tau_{tot}^{-1}$) is the sum of phonon scattering rate ($\tau_{ph}^{-1}$), isotopic scattering rate ($\tau_{iso}^{-1}$), boundary scattering rate ($\tau_b^{-1}$), and strain gradient-induced scattering rate ($\tau_{sg}^{-1}$):

$$\tau_{tot}^{-1} = \tau_{ph}^{-1} + \tau_{iso}^{-1} + \tau_b^{-1} + \tau_{sg}^{-1}. \tag{7}$$

The intrinsic phonon scattering ($\tau_i^{-1}$) includes $\tau_{ph}^{-1}$, $\tau_{iso}^{-1}$, and $\tau_{sg}^{-1}$. As part of intrinsic phonon scattering, the $\tau_{sg}^{-1}$ can be quantized and expressed as [20]:

$$\tau_{sg}^{-1} = \alpha \times f \times \Delta\Theta, \tag{8}$$

where $f$ is the phonon Bose-Einstein distribution at a given temperature, $\Delta\Theta$ represents the phonon frequency shift induced by the strain gradient, and $\alpha$ denotes the ratio of the lattice vibration distance to the characteristic size ($W$) of the inhomogeneous strain field. The term $\alpha \times \Delta\Theta$ characterizes the phonon frequency perturbation. The vibration distance can be evaluated by calculating the atomic interaction cut-off distance $r_c$. Thus, $\alpha$ can be expressed as:

$$\alpha = \frac{r_c}{W}. \tag{9}$$

According to Fermi's golden rules, $\Delta\Theta$ corresponds to the transition rate between phonon mode $\boldsymbol{q}j$ and $\boldsymbol{q}'j'$ and can be computed as [57–61]:

$$\Delta\Theta = \sum_{\boldsymbol{q}'j'} \frac{\pi\Omega}{\omega \boldsymbol{v}_\lambda} |\langle \boldsymbol{q}'j'|H(\omega)|\boldsymbol{q}j\rangle|^2 \delta\left(\omega_{\boldsymbol{q}'j'} - \omega_{\boldsymbol{q}j}\right), \tag{10}$$

where $|\boldsymbol{q}\, j\rangle$ represents the phonon eigenstate, $\boldsymbol{q}$ and $j$ correspond to the wave vector and phonon polarization, respectively. This equation ensures energy conservation in the phonon scattering process. Under the weak perturbation of a strain gradient, the transfer matrix $H$ can be simplified using the Born approximation [59,60]:

$$H \approx H_{\text{sg}} = \frac{\phi_{\text{sg}} - \phi_0}{\sqrt{M_i M_j}}, \tag{11}$$

where $H_{\text{sg}}$ is the perturbation Hamiltonian, $M_i$ and $M_j$ denote the atomic masses of Ga and O, respectively. $\Phi_{\text{sg}}$ and $\Phi_0$ represent the harmonic force constants under inhomogeneous strain and strain-free states, respectively.